# FinTech Adoption, Bank Profitability, and Stability: The Mediating Role of Operational Cost Efficiency in Bangladesh's Commercial Banking Sector

**Shabab Sabiq[a,*], Progga Biswas Parama[b]**

[a] Department of Finance and Banking, University of Barishal, Kornokathi, Barishal-8254, Bangladesh. E-mail: shababsabiq46@gmail.com

[b] Department of Marketing, Eden Mohila College, Affiliated with the University of Dhaka, Azimpur, Lalbag, Dhaka-1205, Bangladesh. E-mail: progga.biswas.mkt@gmail.com

*Corresponding author: Shabab Sabiq, Department of Finance and Banking, University of Barishal, Kornokathi, Barishal-8254, Bangladesh. E-mail: shababsabiq46@gmail.com

## Abstract

While financial technology (FinTech) is seen as a contributor to bank performance, the empirical evidence of digitization's ability to translate into improved profit and stability outcomes and efficiency gains, within emerging systems dominated by banks is not conclusive. The direct impact of a composite FinTech Adoption Index (FTI) on the profitability (ROE) and stability (ln(Z-score)) of Bangladesh's commercial banks is explored in this study along with whether operational cost efficiency (CIR) acts as a mediator. The study is conducted using a balanced panel of 30 commercial banks between 2015 and 2024, supplemented by an entropy-weighted index of FinTech as a robustness check, using a two-step System GMM to address dynamic persistence and endogeneity issues. The adoption of FinTech presents no significant relationship with profitability, stability or cost efficiency, while cost efficiency does not significantly transmit the effect of FinTech to either outcome. The alternative measure suggests that adoption is not only significantly negatively related to profitability, but is also associated with a significantly improved (not worse) cost-to-income ratio, which indicates the sensitivity to measurement rather than any one consistent effect on performance. These findings qualify the expected performance benefits of FinTech and have implications for investment sequencing and regulation of digital transformation in emerging banking markets.

Keywords: FinTech Adoption; Bank Profitability; Bank Stability; Mediator-Operational Cost Efficiency; System GMM; Bangladesh.

JEL Classification: G21, G28, G32, O33

## 1. Introduction

The digitalization of financial services has reshaped the way banks provide services, the pricing of risk, and the management of costs, and has made digital infrastructure a competitive factor (Feyen et al., 2021). For commercial banks, the term FinTech adoption refers to the integration of digital transaction channels, mobile platforms, agent platforms, and automated service infrastructure, which is believed to enhance revenue generation, risk management, and efficiency (Wang et al., 2024).

This assumption has grown into a global issue as the use of digital payments has surged in developing economies (Demirgüç-Kunt et al., 2022), and banks and regulators have recognized the importance of digital transformation for resilience and competitiveness (Feyen et al., 2021). A composite FinTech index has been found to have a Granger-causal relationship with profitability outcomes for a large multi-country sample of banks (Zheng et al., 2023), and digital financial inclusion has been linked to with improved solvency positions (Chinoda and Kapingura, 2023) separately. This is not a consistent finding, however: In some markets, the implementation costs for FinTech have been found to negatively affect profitability before efficiency gains offset them (Yang and Masron, 2023; Zhang, 2026); so the performance payoff of digital investing is not automatic or immediate where adoption is still maturing.

Bangladesh is a distinctive context for this question. The financial system is bank-centric, and agent banking, mobile financial services and digital transaction infrastructure have rapidly gained momentum and registered accounts and agent outlets continue to increase (Bangladesh Bank, 2025). Current evidence for Bangladesh is limited and uses only a single indicator—agent-banking outlet counts—as opposed to using a composite indicator for adoption (Robin et al., 2025).

The impact of FinTech adoption (FTI) on bank profitability (ROE) and bank stability (ln(Z-score)) is important as both outcomes are vital in evaluating digital investment, but rarely examined together. The influence of FinTech adoption on profitability and stability can be mediated through operational cost efficiency (CIR), making it a plausible transmission mechanism linking them. This encourages the analysis of the direct impacts that FinTech has on profitability and stability, as well as its influence on cost efficiency and whether cost efficiency mediates each relationship.

This is because three recurring limitations exist: profitability, stability and efficiency are often studied separately in separate frameworks; the available mediation evidence is geographically limited and contradictory; and the contextual research conducted in Bangladesh is based on very limited FinTech proxies and does not formally test efficiency as a transmission mechanism. To the best of our knowledge, no study has simultaneously examined the direct effects of FinTech adoption on profitability and stability while formally testing operational cost efficiency as a mediator in Bangladesh's commercial banking sector. That is why this study investigates the direct impact of a composite FinTech Adoption Index on ROE and on ln(Z-score), its impact on CIR, and whether CIR is a mediator between the FinTech–profitability and FinTech–stability relationships in the context of commercial banks in Bangladesh. It thus provides an integrated framework that connects the use of FinTech, profitability, stability and cost efficiency; a composite index that is not based on just one measure; and results that are relevant to bank managers, Bangladesh Bank, and policymakers interested in whether the digital transformation is beneficial in terms of measurable returns on performance.

## 2. Literature Review

### *2.1 FinTech Adoption and Bank Profitability*

The idea that digital transformation automatically raises bank profits only has qualified support when the collection of empirical evidence is read as a whole. The profitability of banks is documented to be positively associated with the adoption of FinTech. Composite digital-inclusion indices are found to have a positive relationship with the ROE and ROA of listed Chinese banks (Wang et al., 2024), FinTech dimensions explain most of the variance in deposits, loans, and net profit margin for Jordanian banks (Kaddumi et al., 2023); and a sample of 660 banks in 40 developing countries shows that the PCA-based FinTech index unidirectionally Granger-causes the ROE, ROA and NIM (Zheng et al., 2023). But this connection is neither general nor inevitable. Ovenc and Nabiyev (2025) discover that bank–FinTech partnerships boost ROE only for large Turkish banks, whereas the growth of the FinTech industry itself undermines the profitability of large banks by means of disintermediation, which reveals a simultaneous complementarity effect alongside a substitution effect. Ofosu (2025) also finds that ATM banking raises the profitability of Ghanaian banks, whereas mobile banking lowers it once transaction-levy costs are internalized. Robin et al. (2025) demonstrate that expanding agent-banking outlets improves the profitability of Bangladesh banks, whereas indiscriminate expansion of agent numbers erodes it. More importantly, several dynamic-panel studies isolate a negative direct effect: Yang and Masron (2023) and Zhang (2026) both report that digital transformation alone depresses Chinese bank ROE and ROA, attributing this to front-loaded implementation costs, and Elmahdy et al. (2025) find a similar negative direct effect in Egypt. This apparent contradiction is explained by the mechanism, not the direction: Yang and Masron show the negative effect reverses once interacted with digital financial inclusion, and Elmahdy et al. and Zhang both find the sign flips once operational efficiency is introduced as an explicit pathway, suggesting the profitability payoff of FinTech is transmitted rather than direct. This pattern gives reason to test FinTech's association with profitability as a baseline relationship and to investigate if the mechanism through which the relationship is transmitted is responsible for its sign and magnitude.

**H1:** *FinTech adoption (FTI) is positively associated with bank profitability (ROE), controlling for bank-specific and macroeconomic factors.*

### *2.2 FinTech Adoption and Bank Stability*

The theoretical division of the stability literature is similar: competition-stability vs. competition-fragility hypothesis. One stream finds that FinTech strengthens solvency, Afzal et al. (2025) and Karoui (2025) both observe a positive relationship between FinTech and Z-scores for MENA banks, while Chinoda and Kapingura (2023) note that digital financial inclusion improved Z-scores and decreased NPLs for Sub-Saharan Africa banks, and Chand et al. (2025) report that the FinTech-induced competitive discipline reduced risk-taking for banks in Fiji. Additionally, there is a regulatory channel that imposes limits on the amount of risk-taking by returning deposit funding to traditional banks, as argued by Wu et al. (2024) in their paper, which indicates that post-2016 Chinese FinTech regulation reduced risk-taking by traditional banks. There is a second stream that makes the consensus more complex. In a cross-country sample of over 10,000 institutions, Elekdag et al. (2024) show that FinTech growth is generally associated with heightened risk-taking, consistent with competition-fragility, but the relationship turns around in jurisdictions with strict rule of law and macroprudential oversight, indicating that institutional quality, not FinTech itself, is the pivotal moderating condition. Uddin and Barai (2026) reconcile the two streams temporally: FinTech strains short-run balance-sheet stability through transitional costs in emerging and large Asian economies, but when the capitalization and diversification benefits become obvious, the same adoption improves forward-looking Z-scores, suggesting the sign of the relationship depends partially on the estimation horizon. Building on this, Arebo et al. (2025) introduce another qualification, an inverted-U relationship between financial inclusion and stability in Ethiopia: the higher the capital adequacy, the higher the threshold for further inclusion to be destabilizing. Together, this literature supports the notion that the stability effect of FinTech exists, but is subject to regulatory context and time horizon as well as capital buffers, calling for a direct test of the FinTech–ln(Z-score) relationship as a necessary but not sufficient step towards understanding the conflicting claims.

**H2:** *FinTech adoption (FTI) is positively associated with bank stability ln(Z-score), controlling for bank-specific and macroeconomic factors.*

### *2.3 FinTech Adoption and Operational Cost Efficiency*

A body of efficiency-centric research offers the most plausible explanation for the inconsistencies in profitability and stability. Several studies report that FinTech and digitalisation lead to increases in cost and technical efficiency: Chowdhury et al. (2024) for 15 Asian countries; Li et al. (2025) and Lee et al. (2021) for Chinese banks; Ristanović et al. (2025) for emerging-market banks using blockchain; and Liu et al. (2024) for Pakistani banks, where digitalisation also accelerates efficiency convergence for lagging banks. The efficiency-specific version of the productivity paradox remains a constraint on this evidence, however. According to Saroy et al. (2023), digital payments adoption in India improves cost efficiency without impacting technical efficiency – only if it is done in an integrated ecosystem and not as piecemeal channels. Liao (2023) concludes that the old ATM-based technology reduces the efficiency of Taiwanese banks, whereas the new mobile-payment technology enhances it, with the pattern of raw IT investment yielding weak and inconsistent results: a Solow-paradox pattern echoed by Le et al. (2025), who find that Vietnamese digitalization initially lowers efficiency before a U-shaped reversal beyond an investment threshold. Yoon et al. (2023) expand on this on a global scale, revealing that FinTech can reduce the cost-to-income (CIR) ratio in lower-income nations, but increase it in the richest countries, where a point has been reached where diminishing returns to digital investment have kicked in. These findings suggest that the efficiency effect is not automatic but stage- and tech-dependent, which provides the theoretical basis for hypothesizing a negative FinTech–cost-to-income relationship whose strength will vary with adoption maturity.

**H3:** *FinTech adoption (FTI) is negatively associated with the cost-to-income ratio (CIR), indicating improved operational cost efficiency.*

### *2.4 Cost Efficiency as a Mediator between FinTech Adoption and Bank Profitability and Stability*

The two previous themes point towards an efficiency channel: where FinTech's direct effect on profitability or stability is weak, absent, or perversely signed, an efficiency channel frequently accounts for the discrepancy. This is no longer speculative in the provided literature. Sajid et al. (2023) provide the most direct test: in Pakistani and Bangladeshi banks, operating efficiency fully mediates between FinTech and risk-taking, and in the case of India, it partially mediates between FinTech and risk-taking, while in China efficiency fails to mediate at all and FinTech instead shows a direct positive association with risk-taking, demonstrating that mediation is context-bound rather than universal. For a large Chinese sample, Qin and Jing (2025) add to this by documenting that FinTech lowers NPLs via two mechanisms, cost efficiency and asset efficiency, a finding that sits in tension with Sajid et al.'s null result for China and suggests that the mediation pathway of efficiency may vary depending on the risk dimension and estimation approach used. From a profitability perspective, Elmahdy et al. (2025) report that while FinTech's direct impact is negative on the profitability of Egyptian banks, operational efficiency has a significant positive influence as an intermediary, and Zhang (2026) demonstrates the reverse pattern: digital transformation negatively impacts profitability, precisely because it first erodes operational efficiency and increases NPLs. Li et al. (2022) extend this mediation logic to stability, breaking down the risk-reduction effect of FinTech into a direct component and three indirect channels, of which operational efficiency is the largest. In the context of financial stability, Karoui's (2025) full-mediation finding, although it uses a different mediator, provides additional methodological precedence for examining whether an intermediate operational variable rather than FinTech adoption itself is the proximate driver of bank outcomes. As a whole, this evidence justifies testing operational cost efficiency as a formal mediator of the FinTech–profitability and FinTech–stability relationships, and predicts any such mediation is likely to be partial and context-specific rather than universal across all institutional scenarios.

**H4:** *Operational cost efficiency (CIR) mediates the relationship between FinTech adoption (FTI) and bank profitability (ROE).*
**H5:** *Operational cost efficiency (CIR) mediates the relationship between FinTech adoption (FTI) and bank stability ln(Z-score).*

### *2.5 Research Questions*

Based on the identified research gaps, this study addresses the following research questions:

**RQ1**: Does FinTech adoption have a direct effect on bank profitability (ROE), controlling for bank-specific and macroeconomic factors?

**RQ2**: Does FinTech adoption have a direct effect on bank stability ln(Z-score), controlling for bank-specific and macroeconomic factors?

**RQ3**: Does FinTech adoption improve banks' operational cost efficiency (CIR)?

**RQ4**: Does operational cost efficiency mediate the relationship between FinTech adoption and bank profitability (ROE)?

**RQ5**: Does operational cost efficiency mediate the relationship between FinTech adoption and bank stability ln(Z-score)?

### *2.6 Conceptual Framework*

The study's hypothesized structure based on the conceptual framework is shown in Figure 1. The FinTech Adoption Index (FTI), the independent variable, is constructed using PCA from ATM network density, agent-banking outlets, and digital transaction volume, and for robustness, an entropy-weighted index (FTIEWM) is used instead. FTI is hypothesized to directly influence bank profitability (ROE) and financial stability ln(Z-score), and it is expected to influence operational cost efficiency (CIR), which is hypothesized to mediate the influence of FTI on both outcomes. Bank-specific characteristics (CAR, SIZE, LDR, NPLR, EDR) and macroeconomic variables (INF, GDPGR) are also used as controls affecting the dependent variables.

**Figure 1.**
**Conceptual Framework**

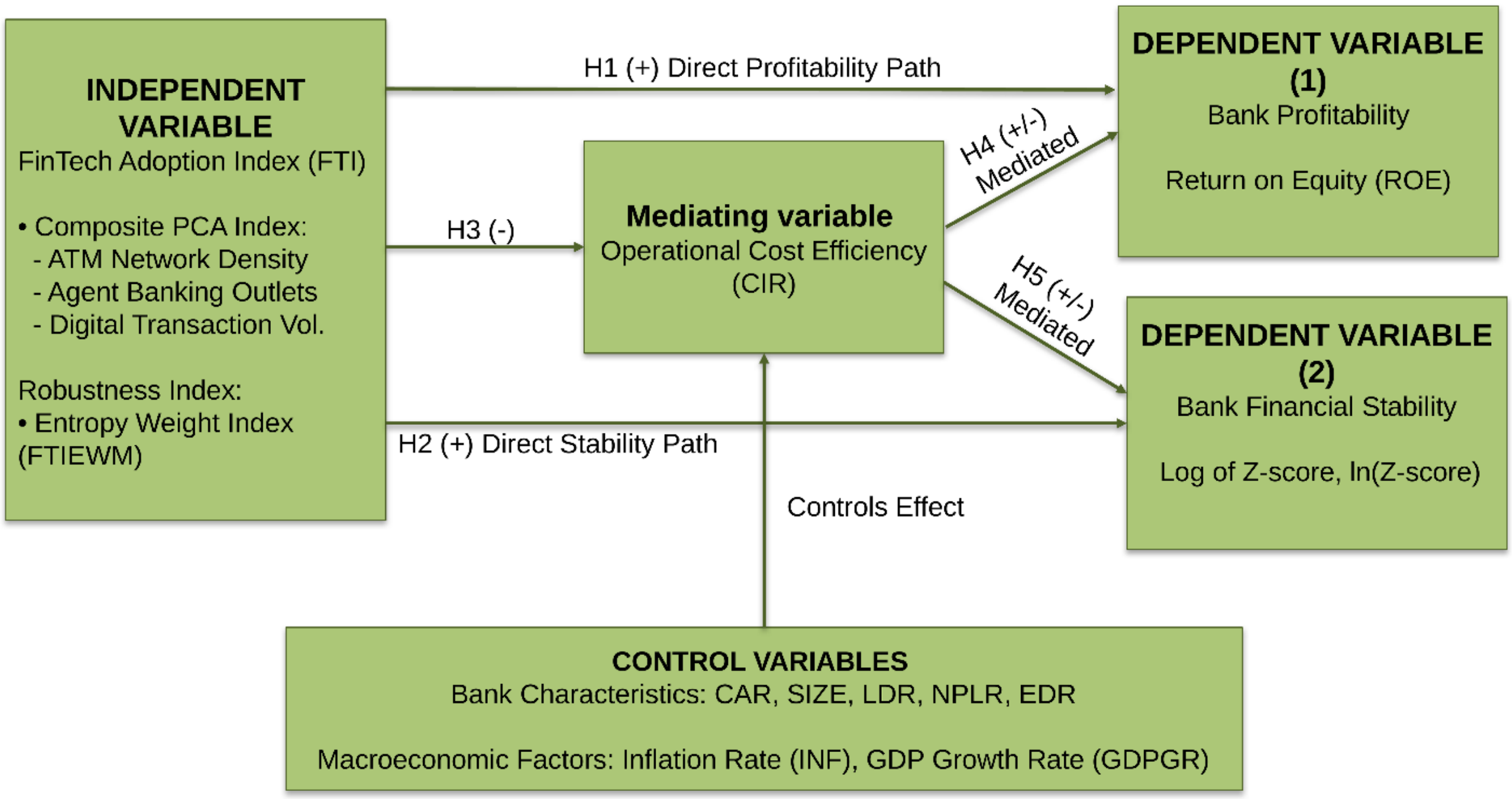

Source: Author's own construction, based on the hypothesized relationships and model specifications developed in this study.

## 3. Methodology

### *3.1 Research Design*

The research design used in this study is a quantitative, panel-based design that aims to investigate the relationship between FinTech adoption and the performance of the banks in a dynamic and heterogeneous institutional environment. Secondary panel data are applied because they can take advantage of the cross-sectional variation across banks and the time-series variation within banks and control for the unobserved time-invariant bank-specific heterogeneity that would have biased estimates in the purely cross-sectional design (Baltagi, 2021). The study population includes commercial banks in Bangladesh, a rapidly digitalizing, bank-dominated emerging financial system in which FinTech adoption varies meaningfully across institutions, offering a suitable empirical setting for testing whether digital transformation translates into measurable profitability, stability, and efficiency outcomes. Following the exclusion of banks with incomplete financial disclosures over the sample window, the final sample comprises a balanced sample of 30 commercial banks observed each year for the period 2015 to 2024, thus, a total of 300 bank-year observations.

### *3.2 Data and Variables*

Data are hand-collected from the audited annual reports of the selected commercial banks, supplemented by Bangladesh Bank publications and banking-sector statistical sources for institutional and industry-level information, while macroeconomic indicators are derived from standard sources of national and international statistics. The central explanatory construct is the FinTech Adoption Index (FTI), which was created by Principal Component Analysis (PCA) from standardized indicators of ATM networks, agent-banking outlets, and volumes of digital transactions (Jolliffe and Cadima, 2016). The index is created by summing each standardized indicator ($f_i$) multiplied by its PCA-derived weight ($w_i$), that is, $FTI = \Sigma_{i=1}^{n}(w_i \times f_i)$, thereby creating a single continuous measure of the intensity of adoption of digital infrastructure rather than relying on a single narrow indicator. For robustness, an alternative index (FTIEWM) is also developed using the Entropy Weight Method, which is based on the information content of each indicator's variation (Zhu et al., 2015), and offers a methodologically different cross-check to the PCA-based measure. There are three outcome variables reflecting bank performance. ROE measures profitability, whereas stability is measured by the natural logarithm of the Z-score (ln(Z-score)), which combines return, volatility, and capitalization into a single inverse measure of insolvency risk (Roy, 1952), and operational cost efficiency is measured by the cost-to-income ratio (CIR), which also serves as the mediating variable between FinTech adoption and profitability and stability. The models also account for the macroeconomic conditions that occurred during the sample period (inflation, INF and GDP growth, GDPGR), and the bank-specific characteristics, namely capital adequacy (CAR), bank size (SIZE), loan-to-deposit ratio (LDR), non-performing loan ratio (NPLR) and equity-to-deposit ratio (EDR). All variables are reported with full definitions, measurement procedures and supporting literature in Table 1.

**Table 1**

**Variable definitions, measurement and sources**

| Nature | Variable | Symbol | Measurement | Source(s) |
|---|---|---|---|---|
| Dependent | Bank Profitability | ROE | Net Income / Average Shareholders' Equity | (Ofosu, 2025; Wang et al., 2024; Yang and Masron, 2023; Zheng et al., 2023) |
| Dependent | Bank Stability | ln(Z-score) | Natural logarithm of Z-score, where Z-score = (ROA + CAR) / Standard Deviation of ROA | (Afzal et al., 2025; Arebo et al., 2025; Elekdag et al., 2024; Zheng et al., 2026) |
| Independent | FinTech Adoption | FTI | PCA-derived weighted FinTech Adoption Index | (Arebo et al., 2025; Zheng et al., 2023; Zheng et al., 2026) |
| Mediator | Cost Efficiency | CIR | Operating Expenses / Operating Income | (Elmahdy et al., 2025; Qin and Jing, 2025; Uddin and Barai, 2026; Zheng et al., 2023) |
| Control | Capital Adequacy Ratio | CAR | (Tier 1 Capital + Tier 2 Capital) / Risk-Weighted Assets | (Abu Khalaf et al., 2025; Arebo et al., 2025; Yang and Masron, 2023; Zheng et al., 2023) |
| Control | Bank Size | SIZE | Natural Logarithm of Total Assets | (Abu Khalaf et al., 2025; Afzal et al., 2025; Ofosu, 2025; Zheng et al., 2023; Zheng et al., 2026) |

| Nature | Variable | Symbol | Measurement | Source(s) |
|---|---|---|---|---|
| Control | Loan-to-Deposit Ratio | LDR | Total Loans / Total Deposits | (Arebo et al., 2025; Wang and Sui, 2021; Yang and Masron, 2023; Zheng et al., 2023) |
| Control | Non-Performing Loan Ratio | NPLR | Total Non-Performing Loans / Total Gross Loans | (Liu et al., 2026; Ofosu, 2025; Qin and Jing, 2025; Zheng et al., 2023) |
| Control | Equity-to-Deposit Ratio | EDR | Total Equity / Total Liabilities | (Wang et al., 2024; Zheng et al., 2023; Zheng et al., 2026) |
| Control | Inflation | INF | Annual Inflation Rate | (Abu Khalaf et al., 2025; Afzal et al., 2025; Zheng et al., 2023; Zheng et al., 2026) |
| Control | GDP Growth | GDPGR | Annual GDP Growth Rate | (Abu Khalaf et al., 2025; Afzal et al., 2025; Zheng et al., 2023; Zheng et al., 2026) |

### 3.3 Model Specification

**Model 1: Profitability Model**

$$ROE_{it} = \alpha + \gamma ROE_{i,t-1} + \beta_1 FTI_{it} + \beta_2 SIZE_{it} + \beta_3 LDR_{it} + \beta_4 NPLR_{it} + \beta_5 CAR_{it} + \beta_6 EDR_{it} + \beta_7 INF_{it} + \beta_8 GDPGR_{it} + \varepsilon_{it} \quad (1)$$

Model 1 regresses ROE on FTI and the control set to test the direct FinTech–profitability relationship (H1)

***Model 2: Stability Model***

$$\ln(ZSCORE)_{it} = \alpha + \gamma \ln(ZSCORE)_{i,t-1} + \beta_1 FTI_{it} + \beta_2 SIZE_{it} + \beta_3 LDR_{it} + \beta_4 NPLR_{it} + \beta_5 CAR_{it} + \beta_6 EDR_{it} + \beta_7 INF_{it} + \beta_8 GDPGR_{it} + \varepsilon_{it} \quad (2)$$

Model 2 applies the same structure to ln(Z-score) to test the direct FinTech–stability relationship (H2).

***Model 3: Cost-Efficiency Model***

$$CIR_{it} = \alpha + \rho CIR_{i,t-1} + \beta_1 FTI_{it} + \beta_2 SIZE_{it} + \beta_3 LDR_{it} + \beta_4 NPLR_{it} + \beta_5 CAR_{it} + \beta_6 EDR_{it} + \beta_7 INF_{it} + \beta_8 GDPGR_{it} + \varepsilon_{it} \quad (3)$$

Model 3 regresses CIR on FTI to establish the first-stage pathway through which FinTech adoption may alter operating costs (H3), a necessary precondition for mediation.

***Model 4: Mediated Profitability Model***

$$ROE_{it} = \alpha + \gamma ROE_{i,t-1} + \beta_1 FTI_{it} + \beta_2 CIR_{it} + \beta_3 SIZE_{it} + \beta_4 LDR_{it} + \beta_5 NPLR_{it} + \beta_6 CAR_{it} + \beta_7 EDR_{it} + \beta_8 INF_{it} + \beta_9 GDPGR_{it} + \varepsilon_{it} \quad (4)$$

Model 4 reintroduces CIR alongside FTI into the ROE equation to formally test whether cost efficiency transmits FinTech's effect onto profitability (H4).

***Model 5: Mediated Stability Model***

$$\ln(ZSCORE_{it}) = \alpha + \gamma \ln(ZSCORE_{i,t-1}) + \beta_1 FTI_{it} + \beta_2 CIR_{it} + \beta_3 SIZE_{it} + \beta_4 LDR_{it} + \beta_5 NPLR_{it} + \beta_6 CAR_{it} + \beta_7 EDR_{it} + \beta_8 INF_{it} + \beta_9 GDPGR_{it} + \varepsilon_{it} \quad (5)$$

Model 5 reintroduces CIR alongside FTI into the ln(Z-score) equation to formally test whether cost efficiency transmits FinTech's effect onto stability (H5). Each model includes the lagged dependent variable to accommodate the persistence documented in bank performance.

### *3.4 Estimation Strategy*

Descriptive statistics, a Pearson correlation matrix and the Variance Inflation Factor are explored before the estimation process and for the VIF values below 10 there is no serious concern of multicollinearity (O'Brien, 2007; Table A.1). Since each model contains a lagged dependent variable, static estimators (pooled OLS, Random Effects, and Fixed Effects) are not applied because the lagged dependent variable is mechanically correlated with the error term in a static framework, and thus model selection procedures like the Hausman test are not done. The panel is instead tested for the Modified Wald test of groupwise heteroskedasticity (Greene, 2012), the Wooldridge (2010) test of serial correlation and the Pesaran (2004) test of cross-sectional dependence (Table A.2). The four reasons for choosing Two-step System GMM (Arellano and Bover, 1995; Blundell and Bond, 1998) as the primary estimator for all five models are: (1) the lagged dependent variable would induce Nickell (1981) bias under a static estimator given the short panel (T = 10); (2) FTI is potentially endogenous, as better-performing banks may be more likely to adopt digital infrastructure; (3) system GMM corrects for both unobserved heterogeneity and dynamic bias together through instrumented differencing; (4) its Windmeijer-corrected standard errors remain valid under the heteroskedasticity, serial correlation, and cross-sectional dependence identified in most models (Table A.2), a fit reinforced by the small-T, large-N structure of the panel (30 banks, 10 years), for which dynamic GMM estimators were developed (Arellano and Bond, 1991), and by the high persistence of the dependent variables. The model is estimated jointly in first differences, which eliminate time-invariant bank effects, and in levels instrumented by lagged differences, which restores efficiency lost through differencing, and the standard errors are corrected as in Windmeijer (2005). The instrument matrix is collapsed with restricted lag depth to reduce instrument proliferation (Roodman, 2009). The tests of the residual serial correlation (AR(1)/AR(2)), the overidentifying restrictions (Sargan and Hansen), and the instrument-subset validity test (Difference-in-Hansen) are used to assess post-estimation validity, with non-rejection providing no evidence against instrument validity rather than confirming it outright. The mediating role of CIR is assessed by comparing FTI's coefficient for the direct models (model 1 and model 2) with the mediated models (models 4 and 5), as well as the significance of the coefficient of CIR in the mediated models. For robustness of the results, all models are re-estimated using FTIEWM in place of FTI. Data is analyzed descriptively in IBM SPSS Statistics and for estimation in STATA 17.0.

## 4. Results

*4.1 Descriptive Statistics*

**Table 2**

**Descriptive Statistics of Study Variables**

| Variable | Obs. | Mean | Std. Dev. | Min | Max |
|---|---|---|---|---|---|
| Return on Equity (ROE) | 300 | 0.1036 | 0.0400 | -0.0346 | 0.2520 |
| Natural Logarithm of Z-score ln(Z-score) | 300 | 3.0789 | 0.8873 | 0.0340 | 4.8355 |
| Cost-to-Income Ratio (CIR) | 300 | 0.4954 | 0.1087 | 0.1596 | 0.8718 |
| FinTech Adoption Index (FTI) | 300 | -0.1455 | 0.3592 | -0.6780 | 1.4846 |
| Bank Size (SIZE) | 300 | 26.4526 | 0.4998 | 25.3185 | 28.4311 |
| Loan-to-Deposit Ratio (LDR) | 300 | 0.7079 | 0.0494 | 0.5013 | 0.8236 |
| Non-Performing Loan Ratio (NPLR) | 300 | 0.0549 | 0.0188 | 0.0208 | 0.1476 |
| Capital Adequacy Ratio (CAR) | 300 | 0.1338 | 0.0181 | 0.0770 | 0.2042 |
| Equity-to-Debt Ratio (EDR) | 300 | 0.0763 | 0.0177 | 0.0405 | 0.1400 |
| Inflation Rate (INF) | 300 | 0.0678 | 0.0181 | 0.0551 | 0.1047 |
| GDP Growth Rate (GDPGR) | 300 | 0.0610 | 0.0128 | 0.0339 | 0.0759 |

Note: Obs. = number of observations; Std. Dev. = standard deviation. All variables are based on 300 bank-year observations covering 30 commercial banks over 2015–2024.

Table 2 provides a summary of the distributional characteristics of the panel for 300 observations (30 banks over 2015-2024). Bank profitability (ROE) averages 10.36% (SD = 0.0400), ranging from a loss of −3.46% to a peak of 25.20%. The stability measure (ln(Z-score)) is on average 3.0789 (SD = 0.8873) and the minimum value is 0.0340, meaning that even the most unstable bank-year observation is positively distant from insolvency. The FinTech Adoption Index (FTI) averages −0.1455 (SD = 0.3592), ranging from −0.6780 to 1.4846, which is in line with a PCA-derived composite rather than a bounded ratio. The cost efficiency (CIR) is on average 49.54% (SD = 0.1087), within a 15.96%–87.18% range typical of emerging-market commercial banking.

*4.2 Correlation Analysis*

**Table 3**

**Pearson Correlation Matrix**

| Variable | ROE | ln(Z-score) | CIR | FTI | SIZE | LDR | NPLR | CAR | EDR | INF | GDPGR |
|---|---|---|---|---|---|---|---|---|---|---|---|
| ROE | 1.0000 | | | | | | | | | | |
| ln(Z-score) | 0.2023*** | 1.0000 | | | | | | | | | |
| CIR | -0.0716 | 0.0055 | 1.0000 | | | | | | | | |
| FTI | -0.0282 | -0.1039* | 0.1587*** | 1.0000 | | | | | | | |
| SIZE | 0.0189 | -0.2667*** | 0.1594*** | 0.4015*** | 1.0000 | | | | | | |
| LDR | -0.1552*** | -0.3063*** | 0.0244 | -0.2015*** | 0.1222** | 1.0000 | | | | | |
| NPLR | -0.3891*** | -0.1115* | 0.0319 | -0.1256** | -0.0369 | 0.0809 | 1.0000 | | | | |
| CAR | 0.1819*** | 0.2330*** | -0.0299 | 0.2523*** | 0.1842*** | -0.3890*** | -0.2480*** | 1.0000 | | | |
| EDR | 0.1260** | 0.1560*** | -0.0546 | -0.2393*** | -0.1801*** | -0.2311*** | -0.0373 | 0.1199** | 1.0000 | | |
| INF | -0.1170** | -0.0891 | -0.0133 | 0.3252*** | 0.4582*** | -0.0712 | 0.1899*** | 0.1599*** | -0.3437*** | 1.0000 | |
| GDPGR | 0.1712*** | 0.0490 | -0.0546 | -0.1987*** | -0.2651*** | 0.1482** | 0.0060 | -0.1133** | 0.1685*** | -0.4430*** | 1.0000 |

Note: Pearson pairwise correlation coefficients are reported. ***, **, and * denote significance at the 1%, 5%, and 10% levels, respectively. Significance levels are based on two-tailed tests.

To screen for multicollinearity before multivariate estimation, pairwise Pearson correlations are reported in Table 3. FTI correlates positively and significantly with SIZE ($r = 0.4015$, $p < 0.01$), CAR ($r = 0.2523$, $p < 0.01$), and INF ($r = 0.3252$, $p < 0.01$). FTI's correlation with CIR is positive but modest ($r = 0.1587$, $p < 0.01$), while correlations with ROE ($r = −0.0282$) and ln(Z-score) ($r = −0.1039$) are weak, foreshadowing the limited direct associations in the multivariate results. The largest coefficient, between SIZE and INF ($r = 0.4582$, $p < 0.01$), remains well below the 0.80 multicollinearity threshold, and the VIF diagnostics (Table A.1), with the highest value (INF, VIF = 1.84) and mean VIF (1.40), are also far below 10

*4.3 Main Dynamic System GMM Regression Results*

**Table 4**

**Main Regression Results: Two-Step System GMM Estimates**

| Variable | Model 1: ROE | Model 2: ln(Z-score) | Model 3: CIR | Model 4: ROE, Mediated | Model 5: ln(Z-score), Mediated |
|---|---|---|---|---|---|
| L.ROE / L.ln(Z-score) / L.CIR | 0.7637*** | 0.7750*** | 0.7092*** | 0.6091** | 0.6194*** |
| | (5.43) | (4.73) | (2.98) | (2.66) | (6.27) |
| FTI | 0.0023 | 0.0328 | -0.0206 | 0.0008 | -0.0080 |
| | (0.07) | (0.65) | (-0.95) | (0.05) | (-0.12) |
| CIR | — | — | — | 0.0668 | 0.0737 |
| | | | | (0.79) | (0.42) |
| SIZE | -0.0009 | -0.1327 | 0.0110 | 0.0037 | -0.2051** |
| | (-0.09) | (-1.27) | (0.61) | (0.27) | (-2.16) |
| LDR | 0.0057 | -0.8243 | 0.1716 | -0.0659 | -1.4435** |
| | (0.08) | (-1.22) | (0.89) | (-0.99) | (-2.69) |
| NPLR | -0.1794 | -1.3477 | -0.3291 | -0.2588 | -2.0016 |
| | (-1.45) | (-1.06) | (-1.14) | (-1.18) | (-0.92) |
| CAR | 0.4095 | 1.5903 | 0.2430 | 0.1676 | 2.0850 |
| | (1.26) | (1.28) | (0.83) | (0.53) | (1.59) |
| EDR | 0.0497 | 1.4016 | 0.1135 | 0.1174 | 0.8544 |
| | (0.14) | (1.04) | (0.24) | (0.52) | (0.45) |
| INF | -0.1153 | 0.4440 | -0.5255 | -0.0223 | 0.7971 |
| | (-0.94) | (0.57) | (-1.29) | (-0.19) | (1.25) |
| GDPGR | 0.4002*** | 0.9391 | -1.2326*** | 0.5248** | 1.3569* |
| | (2.98) | (1.31) | (-3.02) | (2.41) | (1.89) |
| Constant | -0.0250 | 4.4237 | -0.1827 | -0.0926 | 7.1725** |
| | (-0.12) | (1.28) | (-0.49) | (-0.31) | (2.57) |
| Observations | 270 | 270 | 270 | 270 | 270 |
| Number of banks | 30 | 30 | 30 | 30 | 30 |
| Number of instruments | 14 | 14 | 12 | 14 | 17 |
| F-statistic | 777.24*** | 6463.17*** | 1800.39*** | 277.29*** | 1795.55*** |

| Variable | Model 1: ROE | Model 2: ln(Z-score) | Model 3: CIR | Model 4: ROE, Mediated | Model 5: ln(Z-score), Mediated |
|---|---|---|---|---|---|
| AR(1), p-value | 0.010 | 0.013 | 0.039 | 0.040 | 0.018 |
| AR(2), p-value | 0.581 | 0.874 | 0.995 | 0.770 | 0.817 |
| Sargan, p-value | 0.216 | 0.808 | 0.225 | 0.032 | 0.154 |
| Hansen, p-value | 0.139 | 0.821 | 0.237 | 0.173 | 0.797 |
| Difference-in-Hansen, p-value | 0.092 | 0.485 | 0.237 | 0.173 | 0.427 |

Note: Two-step system GMM estimates are reported with Windmeijer-corrected robust standard errors; t-statistics are reported in parentheses. ***, **, and * denote significance at the 1%, 5%, and 10% levels, respectively.

Table 4 reports the two-step System GMM estimates, which are the main ones used to test H1 through H5, with FTI (and CIR in Models 4–5) and the lagged dependent variable instrumented as GMM-style lagged levels/differences, and SIZE, LDR, NPLR, CAR, EDR, INF, and GDPGR as exogenous instruments. The number of instruments (12–17) is still much lower than the number of cross-sectional units, p-values from the Hansen test do not reject the validity of the instruments in each model ($p = 0.139–0.821$), and the AR(2) is not significant at any level ($p \geq 0.581$) in every model. The lagged dependent variable is positive and significant for all five models ($\gamma/\rho = 0.609–0.775$, $p < 0.05$), indicating large persistence. FTI is statistically insignificant for ROE ($\beta = 0.0023$, $t = 0.07$), ln(Z-score) ($\beta = 0.0328$, $t = 0.65$), and CIR ($\beta = -0.0206$, $t = -0.95$); H1, H2, and H3 are not supported. CIR also has an insignificant coefficient in both mediated models, $\beta = 0.0668$, $t = 0.79$ and $\beta = 0.0737$, $t = 0.42$, and in both cases, the coefficient of FTI is also negligible, so neither condition for mediation is met; H4 and H5 are not supported. Among controls, GDPGR is the most consistently significant predictor, raising ROE ($\beta = 0.4002$, $p < 0.01$) and lowering CIR ($\beta = -1.2326$, $p < 0.01$). Models 1, 2 and 5 use GMM-style instruments collapsed over lags 1–2, while Models 3 and 4 use a single collapsed lag (lag 2) to limit instrument proliferation. With one instrument block, the Difference-in-Hansen test has zero degrees of freedom (chi2(0)), so its statistic mechanically equals the overall Hansen statistic in these two models; this is a known xtabond2 property, not a transcription error. To guard against the possible bias of the dynamic-panel specification, the GMM lagged-dependent-variable coefficient lies between the Fixed Effects and pooled OLS estimates in Models 2–5 (Table A.3; Bond, 2002; Flannery and Hankins, 2013), and is 0.011 (0.08 standard errors) higher than the OLS value in Model 1. Adding year dummies to Model 1 leaves FTI insignificant ($\beta = -0.0110$, $t = -0.50$; main estimate $\beta = 0.0023$), with a clean AR(2) test ($p = 0.344$) and no Hansen rejection ($p = 0.266$). Four year dummies were excluded owing to collinearity issues and Stata warns of a singular weighting matrix, so the Sargan rejection ($p = 0.007$) is less informative than the more robust Hansen test. In Model 1, treating CAR, LDR, NPLR and EDR as predetermined variables, rather than exogenous, also leaves FTI insignificant ($\beta = -0.0036$, $t = -0.14$), but this specification is less well identified: 22 instruments, a Sargan rejection ($p < 0.001$), and a Difference-in-Hansen rejection for the remaining exogenous instruments (p = 0.034). This supports keeping the exogenous treatment in the main models.

*4.4 Robustness Analysis Using the Alternative FinTech Adoption Index (EWM)*

**Table 5**

**Robustness Test: System GMM Using the EWM-Based FinTech Index**

| Variable | Model 1: ROE | Model 2: ln(Z-score) | Model 3: CIR | Model 4: ROE, Mediated | Model 5: ln(Z-score), Mediated |
|---|---|---|---|---|---|
| L.ROE / L.ln(Z-score) / L.CIR | 0.7411*** | 0.8651*** | 0.9331*** | 0.6726*** | 0.6223*** |
| | (5.82) | (4.96) | (3.43) | (3.28) | (5.45) |
| FTIEWM | -0.0570** | 0.1445 | -0.9209* | 0.1041 | 0.2634 |
| | (-2.08) | (0.60) | (-1.91) | (0.78) | (0.51) |
| CIR | — | — | — | 0.0518 | 0.0664 |
| | | | | (0.63) | (0.24) |
| SIZE | 0.0063 | -0.0710 | -0.0006 | 0.0024 | -0.1982** |
| | (0.89) | (-0.73) | (-0.03) | (0.21) | (-2.10) |
| LDR | -0.0411 | -0.5240 | 0.0297 | -0.0165 | -1.6452** |
| | (-0.88) | (-0.58) | (0.22) | (-0.17) | (-2.42) |
| NPLR | -0.1238 | -1.0517 | -0.6538 | -0.2084 | -2.1179 |
| | (-0.97) | (-0.83) | (-1.36) | (-1.03) | (-0.86) |
| CAR | 0.3756 | 0.7865 | 0.1456 | 0.1378 | 1.4979 |
| | (1.64) | (0.59) | (0.50) | (0.42) | (0.91) |
| EDR | 0.1115 | 0.9551 | 0.8687 | 0.0551 | 1.1104 |
| | (0.67) | (0.68) | (1.02) | (0.20) | (0.43) |
| INF | -0.1245 | 0.2256 | 0.0072 | -0.0785 | 0.8685 |
| | (-1.54) | (0.28) | (0.01) | (-0.76) | (1.23) |
| GDPGR | 0.5168*** | 0.8204 | -1.2647*** | 0.4850** | 1.3217 |
| | (4.36) | (1.00) | (-3.09) | (2.31) | (1.46) |
| Constant | -0.1861 | 2.4423 | 0.0828 | -0.0836 | 7.2023** |
| | (-1.13) | (0.70) | (0.19) | (-0.34) | (2.54) |
| Observations | 270 | 270 | 270 | 270 | 270 |
| Number of banks | 30 | 30 | 30 | 30 | 30 |
| Number of instruments | 14 | 14 | 12 | 14 | 17 |
| F-statistic | 1236.88*** | 3912.26*** | 2071.59*** | 242.82*** | 1143.53*** |
| AR(1), p-value | 0.004 | 0.014 | 0.047 | 0.026 | 0.021 |
| AR(2), p-value | 0.476 | 0.999 | 0.972 | 0.828 | 0.815 |
| Sargan, p-value | 0.257 | 0.582 | 0.298 | 0.026 | 0.039 |
| Hansen, p-value | 0.468 | 0.387 | 0.219 | 0.123 | 0.355 |
| Difference-in-Hansen, p-value | 0.629 | 0.887 | 0.219 | 0.123 | 0.128 |

Note: Two-step system GMM estimates are reported with Windmeijer-corrected robust standard errors; t-statistics are in parentheses. FTIEWM denotes the FinTech Adoption Index constructed using the Entropy Weight Method. ***, **, and * denote significance at the 1%, 5%, and 10% levels, respectively. Significance levels are based on two-tailed tests.

Table 5 replaces FTI with the entropy-weighted FTIEWM, following the same instrument treatment as Table 4, to assess whether the null findings are measurement-specific. Results are only partially consistent: FTIEWM remains insignificant for ln(Z-score) ($\beta = 0.1445$, $t = 0.60$) and both mediated models ($\beta = 0.1041$, $t = 0.78$; $\beta = 0.2634$, $t = 0.51$), reproducing the null pattern under FTI. However, FTIEWM is significantly negatively associated with ROE ($\beta = -0.0570$, $t = -2.08$, $p < 0.05$) and with CIR ($\beta = -0.9209$, $t = -1.91$, $p < 0.10$); because a lower CIR denotes a lower cost-to-income ratio, the latter indicates improved, not deteriorating, cost efficiency. This diverges from the main estimates, indicating that the profitability and efficiency findings are more sensitive to the measurement of FinTech than the stability and mediation results, which repeat reliably. Diagnostics remain acceptable throughout (Hansen $p = 0.123$–$0.468$; AR(2) $p \geq 0.476$). Again, models 3 and 4 have the same Hansen and Difference-in-Hansen values for the reason given under Table 4.

## 5. Discussion of Findings

### *5.1 FinTech Adoption and Bank Profitability*

On profitability, the System GMM suggests that there is no statistically significant relationship between the intensity of FinTech adoption and profitability ($\beta = 0.0023$, $t = 0.07$) and thus does not support H1. This is in contrast with a finding from more digitally advanced systems that composite FinTech indices generate higher returns on equity and on assets (Wang et al., 2024; Zheng et al., 2023), but consistent with studies attributing early-stage digital investment to front-loaded implementation costs that offset earnings gains (Elmahdy et al., 2025; Yang and Masron, 2023; Zhang, 2026). According to Transaction Cost Theory, the cost benefits digitalisation offers are yet to be significant to shift profit levels; Bangladesh's commercial banks are still in an adoption phase before they can achieve measurable financial benefits, according to Diffusion of Innovation Theory. The null finding is not completely confirmed by the entropy-weighted robustness check: FTIEWM is found to be significantly negatively associated with ROE ($\beta = -0.0570$, $t = -2.08$, $p < 0.05$), so the absence of a profitability effect should not be treated as fully settled.

### *5.2 FinTech Adoption and Bank Stability*

There is a positive relationship though not statistically significant, between the use of FinTech and stability ($\beta = 0.0328$, $t = 0.65$), which does not support H2. This falls between the competition-stability perspective in which digital tools strengthen monitoring and credit assessment (Afzal et al., 2025; Chinoda and Kapingura, 2023), and the competition-fragility perspective in which greater access to digital tools increases risk-taking more rapidly than risk-management capacity can respond (Elekdag et al., 2024). The insignificant result suggests that the Bangladeshi banking sector is not yet mature enough for either of the two forces to have a significant impact, consistent with transitional patterns observed in other Asian banking systems (Uddin and Barai, 2026). This null finding is more stable across measurement choices, as it replicates under FTIEWM ($\beta = 0.1445$, $t = 0.60$), compared to the profitability null finding.

*5.3 FinTech Adoption and Operational Cost Efficiency*

There is no evidence that FinTech adoption has a significant effect on CIR under the main estimates ($\beta = -0.0206$, $t = -0.95$). H3 is therefore not supported. This is in contrast to evidence of benefits from efficiency gains via automation and digital delivery (Chowdhury et al., 2024; Lee et al., 2021), although it broadly aligns with the efficiency-specific productivity paradox, which sees technology costs and benefits cancel each other out and then a reversal (Le et al., 2025; Saroy et al., 2023). The robustness check adds to this picture: FTIEWM is significantly associated with a lower CIR ($\beta = -0.9209$, $t = -1.91$, $p < 0.10$), which suggests better, rather than worse, cost efficiency. It is not obvious how this can be reconciled with the negative impact of FTIEWM on ROE, though, as the lower cost-to-income ratio should have a positive impact on profitability; the savings may be offset elsewhere also (for example, in provisioning) and this cannot be confirmed without breaking out the impact on ROE, and is best read as a measurement-sensitive anomaly. The lagged CIR coefficient also rises from 0.709 (Table 4) to 0.933 (Table 5); both are below one and AR(2) is clean in both ($p = 0.995$, 0.972), so this reflects measurement sensitivity rather than a unit root or misspecification.

*5.4 Cost Efficiency as a Mediator*

Mediation requires FTI to move CIR, and CIR to move profitability or stability. Neither link holds in the main estimates: FTI is insignificant in the Model 3 first stage and CIR in Models 4 ($\beta = 0.0668$, $t = 0.79$) and 5 ($\beta = 0.0737$, $t = 0.42$), and FTI's coefficient remains negligible in the mediated specifications. H4 and H5 are not supported. This null result is replicated under FTIEWM in both mediated models ($\beta = 0.1041$, $t = 0.78$; $\beta = 0.2634$, $t = 0.51$), lending confidence to the conclusion regardless of FinTech measurement. This is contrary to findings that show mediation through operational efficiency, either full or partial (Elmahdy et al., 2025; Qin and Jing, 2025), but consistent with findings that mediation through efficiency is very context-specific itself (Sajid et al., 2023).

## 6. Conclusion

FinTech is often seen as a straightforward route to stronger, more resilient banks, but this presumption overlooks that the three aspects of a bank's success – profitability, safety, and efficiency – need not move together. This difference is important in Bangladesh, where agent banking and digital infrastructure have grown at a fast pace across a diverse set of commercial banks, but where no study has formally tested whether this expansion is paying off, directly or through the efficiency channel it is widely assumed to generate. This study addressed that gap by examining the direct effect of FinTech on profitability (ROE) and stability (ln(Z-score)), its effect on cost efficiency (CIR), and whether cost efficiency transmits FinTech's effect to either outcome, using the dynamic System GMM as the main estimator. The overall results of the main evidence indicate that there is no significant association found between the adoption of FinTech and profitability, stability or cost efficiency, nor do they support either hypothesised mediation pathway. The robustness check qualifies this picture: the stability and mediation findings replicate under the entropy-weighted index, but FTIEWM is significantly related to lower ROE and improved cost efficiency, as CIR is a cost-to-income ratio. That an apparent efficiency gain does not raise profitability suggests any savings are not reaching the bottom line. Taken together, H1 through H5 are not supported by the primary GMM evidence.

*6.1 Theoretical Implications*

These findings qualify rather than overturn Transaction Cost Theory and Diffusion of Innovation Theory. The direct impacts were mostly null, consistent with the early stages of the adoption process, where fixed costs remain high and are likely to be the reason for the cost savings expected by these theories not yet being realised. The same interpretation goes for the lack of mediation: not that it is impossible for operational efficiency to play a conduit role in creating the effects of FinTech, but that this role has yet to be activated in this context. The findings regarding the profitability and efficiency were also sensitive to the measurement used, indicating that the operationalisation of FinTech adoption – rather than just its existence – is important to the results.

### *6.2 Managerial and Policy Implications*

The findings indicate that while the introduction of FinTech solutions is not enough, in isolation, to drive measurable returns for banks, the process probably requires some support from organisational change, employees' capabilities and processes, among other things. To Bangladesh Bank and the regulators, the message is not to slow down FinTech growth but to pair it with closer supervisory attention to adoption quality, as the intensity of adoption is not an accurate measure of better bank performance.

### *6.3 Limitations and Directions for Future Research*

This study is limited in scope by its single-country design, a ten-year panel which may understate effects unfolding over a longer adoption horizon, and by the composite FinTech measure which is cross-validated with an alternative based on the entropy measure of digital indicators but still captures only a small number of digital indicators. System GMM is able to deal with much of the endogeneity in dynamic bank panels, but not all. Future studies may investigate whether these are found in other countries, if the relationship between FinTech and performance is nonlinear and not linear, and if the impact of FinTech differs between state-owned and private banks, or across banks of different sizes.

## Ethical statement

Ethical approval was not required for this study as it did not involve human participants, animal subjects, or the use of identifiable or sensitive data.

## Consent/Human/Animal statement

The authors declare that the work described does not involve patients or volunteers. The authors declare that the work described has not involved experimentation on humans or animals.

## Data availability

Data used in this study are available upon reasonable request to the corresponding author.

## References

Abu Khalaf, B., Al-Sharkas, A., Sarea, A., 2025. Realizing opportunities: The influence of FinTech on the success of MENA banks. *Discov. Sustain.* 6, 501. https://doi.org/10.1007/s43621-025-01374-8.

Afzal, A.M., Abu Khalaf, B., Al-Naimi, M.S., Samara, E., 2025. The impact of Fintech on the stability of Middle Eastern and North African (MENA) banks. *Risks* 13, 106. https://doi.org/10.3390/risks13060106.

Arebo, M., Hando, F., Mekonnen, A., 2025. Financial inclusion and stability in Ethiopia using bank-level data: A two-step system GMM estimation. *F1000Research* 13, 1369. https://doi.org/10.12688/f1000research.158461.2.

Arellano, M., Bond, S., 1991. Some tests of specification for panel data: Monte Carlo evidence and an application to employment equations. *Rev. Econ. Stud.* 58, 277-297. https://doi.org/10.2307/2297968.

Arellano, M., Bover, O., 1995. Another look at the instrumental variable estimation of error-components models. *J. Econom.* 68, 29-51. https://doi.org/10.1016/0304-4076(94)01642-D.

Baltagi, B.H., 2021. *Econometric Analysis of Panel Data*, sixth ed. Springer, Cham. https://doi.org/10.1007/978-3-030-53953-5.

Bangladesh Bank, 2025. *Mobile Financial Services (MFS) Comparative Summary Statement*. Payment Systems Department, Bangladesh Bank. https://www.bb.org.bd/en/index.php/financialactivity/mfsdata.

Blundell, R., Bond, S., 1998. Initial conditions and moment restrictions in dynamic panel data models. *J. Econom.* 87, 115-143. https://doi.org/10.1016/S0304-4076(98)00009-8.

Bond, S.R., 2002. Dynamic panel data models: A guide to micro data methods and practice. *Port. Econ. J.* 1, 141-162. https://doi.org/10.1007/s10258-002-0009-9.

Chand, S.A., Singh, B., Narayan, K., Chand, A., 2025. The impact of financial technology (FinTech) on bank risk-taking and profitability in small developing island states: A study of Fiji. *J. Risk Financ. Manag.* 18, 366. https://doi.org/10.3390/jrfm18070366.

Chinoda, T., Kapingura, F.M., 2023. The impact of digital financial inclusion and bank competition on bank stability in Sub-Saharan Africa. *Economies* 11, 15. https://doi.org/10.3390/economies11010015.

Chowdhury, M.A., Soh, W.N., Abdul Rahim, N.B., Johari, J.B., 2024. How Fintech is doing innovations and efficiency in the banking sector? *Int. J. Bus. Soc.* 25, 888-907. https://doi.org/10.33736/ijbs.8514.2024.

Demirgüç-Kunt, A., Klapper, L., Singer, D., Ansar, S., 2022. *The Global Findex Database 2021: Financial Inclusion, Digital Payments, and Resilience in the Age of COVID-19*. World Bank, Washington, DC.

Elekdag, S., Emrullahu, D., Ben Naceur, S., 2024. Does FinTech increase bank risk taking? IMF Working Paper WP/24/17. International Monetary Fund, Washington, DC. https://www.imf.org/en/Publications/WP/Issues/2024/01/26/Does-FinTech-Increase-Bank-Risk-Taking-544081.

Elmahdy, A.H.A.M., Abdelkader, M.T.K.M., Shaker, M.A.M., 2025. Bridging the nexus between Fintech, operational efficiency and banks profitability: The moderating role of bank size. *Future Bus. J.* 11, 62. https://doi.org/10.1186/s43093-025-00478-x.

Feyen, E., Frost, J., Gambacorta, L., Natarajan, H., Saal, M., 2021. *Fintech and the Digital Transformation of Financial Services: Implications for Market Structure and Public Policy*. BIS Papers No. 117. Bank for International Settlements, Basel.

Flannery, M.J., Hankins, K.W., 2013. Estimating dynamic panel models in corporate finance. *J. Corp. Financ.* 19, 1-19. https://doi.org/10.1016/j.jcorpfin.2012.09.004.

Greene, W.H., 2012. *Econometric Analysis*, seventh ed. Prentice Hall, Boston.

Jolliffe, I.T., Cadima, J., 2016. Principal component analysis: A review and recent developments. *Philos. Trans. R. Soc. A Math. Phys. Eng. Sci.* 374, 20150202. https://doi.org/10.1098/rsta.2015.0202.

Kaddumi, T.A., Baker, H., Nassar, M.D., A-Kilani, Q., 2023. Does financial technology adoption influence bank's financial performance: The case of Jordan. *J. Risk Financ. Manag.* 16, 413. https://doi.org/10.3390/jrfm16090413.

Karoui, C., 2025. Exploring the mediating role of financial stability in the relationship between FinTech and bank performance: Evidence from the MENA region. *J. Ecohumanism* 4, 264-277. https://doi.org/10.62754/joe.v4i3.6623.

Le, T., Ngo, T., Nguyen, D.T., Ho, T.H., Do, T.T.M., 2025. Digitalisation and bank efficiency: Evidence from an emerging market. *Asian Acad. Manag. J. Account. Financ.* 21, 29-65. https://doi.org/10.21315/aamjaf2025.21.2.2.

Lee, C.-C., Li, X., Yu, C.-H., Zhao, J., 2021. Does Fintech innovation improve bank efficiency? Evidence from China's banking industry. *Int. Rev. Econ. Financ.* 74, 468-483. https://doi.org/10.1016/j.iref.2021.03.009.

Li, G., Elahi, E., Zhao, L., 2022. Fintech, bank risk-taking, and risk-warning for commercial banks in the era of digital technology. *Front. Psychol.* 13, 934053. https://doi.org/10.3389/fpsyg.2022.934053.

Li, Y., Stasinakis, C., Yeo, W.M., Da Silva Fernandes, F., 2025. Fintech, financial development and banking efficiency: Evidence from Chinese commercial banks. *Eur. J. Financ.* 31, 1245-1295. https://doi.org/10.1080/1351847X.2025.2468481.

Liao, C.-S., 2023. How does Fintech affect bank efficiency in Taiwan? *PLoS One* 18, e0289629. https://doi.org/10.1371/journal.pone.0289629.

Liu, T.L., Naveed, M.M., Mustafa, S., Naveed, M.T., 2024. Analysing the impact of digital technology diffusion on the efficiency and convergence process of the commercial banking industry of Pakistan. *Humanit. Soc. Sci. Commun.* 11, 673. https://doi.org/10.1057/s41599-024-03184-1.

Liu, Y., Tong, A., Nourani, P., 2026. FinTech adoption and credit risk dynamics: Evidence from China's A-Share banking sector. *J. Bank. Regul.* 27, 25. https://doi.org/10.1057/s41261-026-00327-z.

Nickell, S., 1981. Biases in dynamic models with fixed effects. *Econometrica* 49, 1417-1426. https://doi.org/10.2307/1911408.

Ofosu, P.K., 2025. Financial technology and bank profitability: Ghanaian evidence. *J. Econ. Bus. Manag. Issues* 2, 491-508. https://doi.org/10.47134/jebmi.v2i4.689.

Ovenc, G., Nabiyev, A.B., 2025. Discover how Fintech is transforming bank performance: Insights from an emerging economy. *Cogent Econ. Financ.* 13, 2477676. https://doi.org/10.1080/23322039.2025.2477676.

O'Brien, R.M., 2007. A caution regarding rules of thumb for variance inflation factors. *Qual. Quant.* 41, 673-690. https://doi.org/10.1007/s11135-006-9018-6.

Pesaran, M.H., 2004. General diagnostic tests for cross-section dependence in panels. Cambridge Working Papers in Economics No. 0435. University of Cambridge, Cambridge. https://doi.org/10.17863/CAM.5113.

Qin, Z., Jing, Z., 2025. Fintech adoption and credit risk mitigation: Evidence from Chinese commercial banks. *Sustainability* 17, 10294. https://doi.org/10.3390/su172210294.

Ristanović, V., Primorac, D., Mulović Trgovac, A., 2025. Banking in the age of blockchain and FinTech: A hybrid efficiency framework for emerging economies. *J. Risk Financ. Manag.* 18, 458. https://doi.org/10.3390/jrfm18080458.

Robin, I.A., Islam, M.M., Alharthi, M., 2025. The impact of FinTech on the financial performance of commercial banks in Bangladesh: A random-effect model analysis. *FinTech* 4, 40. https://doi.org/10.3390/fintech4030040.

Roodman, D., 2009. A note on the theme of too many instruments. *Oxf. Bull. Econ. Stat.* 71, 135-158. https://doi.org/10.1111/j.1468-0084.2008.00542.x.

Roy, A.D., 1952. Safety first and the holding of assets. *Econometrica* 20, 431-449. https://doi.org/10.2307/1907413.

Sajid, R., Ayub, H., Malik, B.F., Ellahi, A., 2023. The role of Fintech on bank risk-taking: Mediating role of bank's operating efficiency. *Hum. Behav. Emerg. Technol.* 2023, 7059307. https://doi.org/10.1155/2023/7059307.

Saroy, R., Jain, P., Awasthy, S., Dhal, S.C., 2023. Impact of digital payment adoption on Indian banking sector efficiency. *J. Bank. Financ. Technol.* 7, 47. https://doi.org/10.1007/s42786-023-00047-2.

Uddin, H., Barai, M.K., 2026. Fintech adoption and bank risk, efficiency and stability: Evidence from panel data of selected Asian economies. *FinTech* 5, 14. https://doi.org/10.3390/fintech5010014.

Wang, Y., Sui, X., 2021. Can Fintech improve the efficiency of commercial banks?—An analysis based on big data. *Res. Int. Bus. Financ.* 55, 101338. https://doi.org/10.1016/j.ribaf.2020.101338.

Wang, Y., Yu, X., Yao, Q., Lu, Y., Che, W., Jiang, J., Chen, S.C.-I., 2024. Assessing the impact of financial technology innovations on the sustainable profitability of listed commercial banks in China. *FinTech* 3, 337-348. https://doi.org/10.3390/fintech3030019.

Windmeijer, F., 2005. A finite sample correction for the variance of linear efficient two-step GMM estimators. *J. Econom.* 126, 25-51. https://doi.org/10.1016/j.jeconom.2004.02.005.

Wooldridge, J.M., 2010. *Econometric Analysis of Cross Section and Panel Data*, second ed. MIT Press, Cambridge, MA.

Wu, Z., Li, L., Wang, B., Zhang, X., 2024. FinTech regulation and banks' risk-taking: Evidence from China. *PLoS One* 19, e0311722. https://doi.org/10.1371/journal.pone.0311722.

Yang, F., Masron, T.A., 2023. Does financial inclusion moderate the effect of digital transformation on banks' performance in China? *Cogent Econ. Financ.* 11, 2267270. https://doi.org/10.1080/23322039.2023.2267270.

Yoon, S.S., Lee, H., Oh, I., 2023. Differential impact of Fintech and GDP on bank performance: Global evidence. *J. Risk Financ. Manag.* 16, 304. https://doi.org/10.3390/jrfm16070304.

Zhang, S., 2026. The influence of digital transformation on Chinese bank profitability performance. *PLoS One* 21, e0340249. https://doi.org/10.1371/journal.pone.0340249.

Zheng, C., Rahman, M.A., Hossain, S., Moudud-Ul-Huq, S., 2023. Does Fintech-driven inclusive finance induce bank profitability? Empirical evidence from developing countries. *J. Risk Financ. Manag.* 16, 457. https://doi.org/10.3390/jrfm16100457.

Zheng, C., Rahman, M.A., Hossain, S., Moudud-Ul-Huq, S., 2026. Impact of Fintech-centric financial inclusion on bank risk-taking: Evidence from developing countries. *Financ. Innov.* 12, 58. https://doi.org/10.1186/s40854-025-00843-3.

Zhu, Y., Hipel, K.W., Ge, Y., Cao, M., 2015. Combined weights approach for a multi-attribute group decision analysis methodology for fire risk decision making. *Kybernetes* 44, 837-858. https://doi.org/10.1108/K-04-2014-0079.

## Appendix

**Table A1. Variance Inflation Factor (VIF) Test for Multicollinearity**

| Variable | VIF | 1/VIF (Tolerance) |
|---|---|---|
| INF | 1.84 | 0.543713 |
| SIZE | 1.68 | 0.596315 |
| LDR | 1.49 | 0.670973 |
| FTI | 1.47 | 0.681166 |
| CAR | 1.43 | 0.701300 |
| EDR | 1.32 | 0.756845 |
| GDPGR | 1.30 | 0.769864 |
| Ln(Z-score) | 1.26 | 0.795840 |
| NPLR | 1.19 | 0.839922 |
| CIR | 1.08 | 0.928109 |
| Mean VIF | 1.40 | — |

Note: VIF and 1/VIF (tolerance) values are reported to assess multicollinearity among the explanatory variables. The results indicate no serious multicollinearity concerns, as the mean VIF is 1.40 and all individual VIF values are below 10.

**Table A2**

**Diagnostic Tests for Panel Data Models**

| Diagnostic Test | Model 1 (ROE) | Model 2 ln(Z-score) | Model 3 (CIR) | Model 4 (ROE, Mediated) | Model 5 ln(Z-score, Mediated) |
|---|---|---|---|---|---|
| Modified Wald Test for Groupwise Heteroskedasticity | $\chi^2(30) = 493.78$ | $\chi^2(30) = 1542.74$ | $\chi^2(30) = 33.06$ | $\chi^2(30) = 483.19$ | $\chi^2(30) = 1499.52$ |
| p-value | 0.0000 | 0.0000 | 0.3200 | 0.0000 | 0.0000 |
| Wooldridge Test for Serial Correlation | F(1,29) = 26.029 | F(1,29) = 56.574 | F(1,29) = 20.000 | F(1,29) = 25.867 | F(1,29) = 45.142 |
| p-value | 0.0000 | 0.0000 | 0.0001 | 0.0000 | 0.0000 |
| Pesaran Test for Cross-sectional Dependence | 0.945 | 3.424 | 5.293 | 0.990 | 5.520 |
| p-value | 0.3444 | 0.0006 | 0.0000 | 0.3220 | 0.0000 |
| Average Absolute Off-diagonal Correlation | 0.306 | 0.304 | 0.319 | 0.306 | 0.294 |
| Overall Diagnostic Decision | Heteroskedasticity; Serial correlation; No cross-sectional dependence | Heteroskedasticity; Serial correlation; Cross-sectional dependence | No heteroskedasticity; Serial correlation; Cross-sectional dependence | Heteroskedasticity; Serial correlation; No cross-sectional dependence | Heteroskedasticity; Serial correlation; Cross-sectional dependence |

Note: Decisions are based on the respective test null hypotheses. ***, **, and * denote significance at the 1%, 5%, and 10% levels, respectively.

**Table A3**

**Lagged Dependent Variable: Pooled OLS, Fixed Effects, and System GMM Bounds Check**

| Statistic | Model 1 (ROE) | Model 2 ln(Z-score) | Model 3 (CIR) | Model 4 (ROE, Mediated) | Model 5 ln(Z-score, Mediated) |
|---|---|---|---|---|---|
| Pooled OLS, L.DV coefficient | 0.7526 | 0.9862 | 0.8575 | 0.7529 | 0.9868 |
| Fixed Effects, L.DV coefficient | 0.5193 | 0.4963 | 0.5317 | 0.5164 | 0.4450 |
| System GMM, L.DV coefficient (s.e.) | 0.7637 (0.1406) | 0.7750 (0.1639) | 0.7092 (0.2383) | 0.6091 (0.2288) | 0.6194 (0.0988) |
| GMM within [FE, Pooled OLS] bounds? | Marginal (+0.011 above OLS; 0.08 s.e.) | Yes | Yes | Yes | Yes |

Note: Pooled OLS and Fixed Effects coefficients come from baseline estimators run for comparison only; GMM coefficients (Windmeijer-corrected s.e.) are from Table 4. A consistent estimator should lie between the Fixed Effects (downward-biased) and pooled OLS (upward-biased) values (Bond, 2002; Flannery and Hankins, 2013).